\documentclass[conference]{IEEEtran}
\IEEEoverridecommandlockouts
\usepackage{cite}
\usepackage{amsmath,amssymb,amsfonts}
\usepackage{algorithm}
\usepackage{algorithmic}
\usepackage{graphicx}
\usepackage{textcomp}
\usepackage{xcolor}
\usepackage{multirow}
\usepackage{tikz}
\usetikzlibrary{shapes.geometric, arrows.meta, positioning, fit, backgrounds}
\def\BibTeX{{\rm B\kern-.05em{\sc i\kern-.025em b}\kern-.08em
    T\kern-.1667em\lower.7ex\hbox{E}\kern-.125emX}}
\begin{document}

\title{R$^3$D: Regional-guided Residual Radar Diffusion}

\author{
		Hao Li,University of Arizona\quad
		Xinqi Liu,University of Hong Kong \quad
		Yaoqing Jin,University of Stuttgart\\
		\texttt{lihao@arizona.edu} \quad
		\texttt{Xinqi\_liu28@connect.hku.hk} \quad
		\texttt{st186181@stud.uni-stuttgart.de} 
	}
	
\maketitle

\begin{abstract}
Millimeter-wave radar enables robust environment perception in autonomous systems under adverse conditions yet suffers from sparse, noisy point clouds with low angular resolution. Existing diffusion-based radar enhancement methods either incur high learning complexity by modeling full LiDAR distributions or fail to prioritize critical structures due to uniform regional processing. To address these issues, we propose R³D, a regional-guided residual radar diffusion framework that integrates residual diffusion modeling—focusing on the concentrated LiDAR-radar residual encoding complementary high-frequency details to reduce learning difficulty—and $\sigma$-adaptive regional guidance—leveraging radar-specific signal properties to generate attention maps and applying lightweight guidance only in low-noise stages to avoid gradient imbalance while refining key regions. Extensive experiments on the ColoRadar dataset demonstrate that R$^3$D outperforms state-of-the-art methods, providing a practical solution for radar perception enhancement. Our anonymous code and pretrained models are released here: https://anonymous.4open.science/r/r3d-F836

\end{abstract}

\begin{IEEEkeywords}
mmWave radar; point cloud enhancement; diffusion model; residual learning; regional guidance
\end{IEEEkeywords}

\section{Introduction}\label{intro}
Millimeter wave (mmWave) radars have become indispensable sensing modalities for autonomous systems (e.g., micro aerial vehicles (MAVs) and advanced driver assistance systems (ADAS)) due to their robustness against extreme weather conditions such as rain, fog, and low light. However, inherent limitations including poor angular resolution, sparse signal distribution, and severe noise interference result in radar point clouds that are far less dense and accurate than LiDAR point clouds. This deficiency hinders their deployment in cluttered environments requiring precise simultaneous localization and mapping (SLAM) and environmental perception.

To address the sparsity and noise issues of mmWave radar data, recent studies have explored cross-modal learning frameworks, where LiDAR point clouds serve as ground truth to guide radar point cloud enhancement. Among these methods, diffusion models have emerged as powerful generative tools, achieving state-of-the-art performance in radar point cloud generation by modeling the mapping from radar range-azimuth heatmaps (RAHs) to LiDAR bird’s eye view (BEV) images. Despite their success, existing radar diffusion methods still face two critical challenges that limit their practicality and performance.
\begin{figure}[t!]
	\centering
	\label{fig:residual_vs_lidar}
	\vspace{5pt}
	\includegraphics[width=0.98\linewidth]{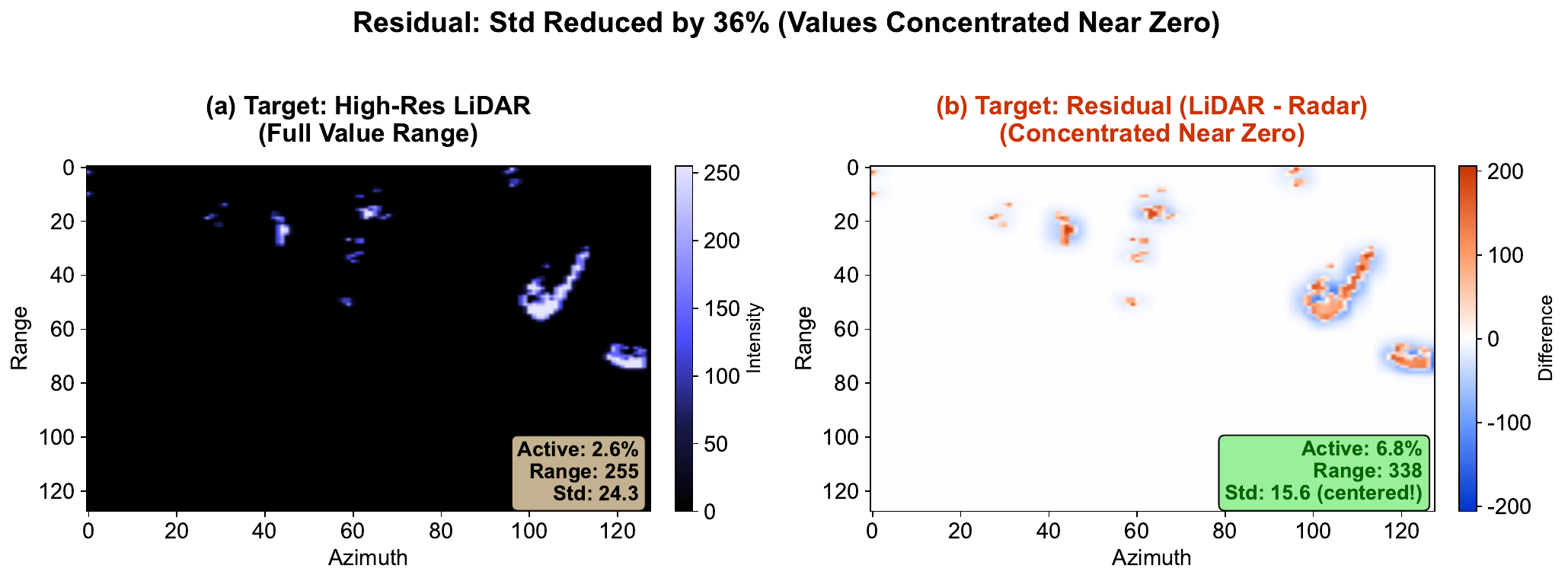}
	\vspace{3pt}
	\caption{Residual Learning Concentrates Target Distribution Near Zero. Left: High-Res LiDAR target (Active: 2.6\%, Range: 255, full value distribution); Right: Residual (LiDAR - Radar) target (Active: 6.8\%, Range: 338 (absolute range), Std: 15.6, centered near zero). Although the residual has a larger absolute range, 93.2\% of its valid values are within ±10 (highly concentrated around zero), resulting in a significantly smaller standard deviation. This concentration makes residual learning substantially easier than modeling the full LiDAR distribution, as the model only needs to predict small deviations from zero rather than the entire value spectrum.}
\end{figure}

First, existing radar diffusion models directly model the full distribution of LiDAR point clouds conditioned on radar data, leading to high learning complexity. Such approaches require the model to learn all spatial features of LiDAR, including redundant information that overlaps with radar signals. This not only increases computational overhead but also dilutes the model’s focus on high-frequency details (e.g., object edges and structural boundaries) that radar inherently lacks. In the super-resolution field, it is widely recognized that recovering the residual (i.e., the difference between high-resolution and low-resolution data) is substantially easier(as Fig.\ref{fig:residual_vs_lidar} shows) than generating high-resolution content from scratch—this is because residuals primarily encode complementary high-frequency details rather than redundant base information. Yet this well-established paradigm has not been extended to radar diffusion for point cloud enhancement.

Second, existing diffusion-driven radar enhancement techniques adopt a uniform processing strategy for all regions of the input during denoising. Radar RAHs exhibit distinct spatial heterogeneity: strong scattering regions (e.g., buildings, vehicles) contain critical environmental structure information, while weak scattering regions are dominated by noise and clutter. While uniform processing is necessary for learning global structure in early denoising stages, indiscriminately applying the same optimization emphasis to all regions throughout the entire diffusion process wastes computational resources on noise-dominated areas and may fail to adequately refine key structural regions during detail recovery phases.

Although attention-guided diffusion methods have been proposed for image super-resolution to address this issue, they rely on natural image-oriented self-supervised frameworks (e.g., DINO) that suffer from domain mismatch when applied to radar signals, and their complex pre-training requirements are incompatible with the limited availability of radar datasets. In stark contrast, radar signal properties—such as intensity and local consistency—can directly identify target regions without domain adaptation or pre-training. However, naively applying aggressive regional weighting throughout the entire diffusion process disrupts the gradient balance of noise-level-dependent loss weighting (e.g., Karras weighting $w(\sigma_t) = 1/\sigma_t^2$), leading to training instability. Therefore, a principled approach is needed to harmonize regional guidance with diffusion models' inherent optimization dynamics.

Based on these, we propose R$^3$D--Regional guided Residual Radar Diffusion-- a novel cross-modal diffusion framework that integrates residual learning and $\sigma$-adaptive regional guidance for dense and accurate radar point cloud generation. The core contributions of this work are twofold:
\begin{itemize}
	\item \textbf{Residual Diffusion Modeling}: We model the residual distribution between LiDAR and radar signals. This residual explicitly encodes the high-frequency details and structural information that radar lacks, reducing the learning complexity by focusing on the "complementary information" rather than the full LiDAR distribution. 
	\item \textbf{$\sigma$-Adaptive Regional Guidance}: By leveraging radar signal properties to generate attention maps without pre-training, our method applies lightweight regional weighting only during low-noise detail recovery stages, while maintaining uniform processing during high-noise global structure learning stages.
\end{itemize}

We validate R³D on the ColoRadar dataset encompassing seven diverse scenes. Experimental results demonstrate that R³D outperforms state-of-the-art radar enhancement methods: Proposed-Residual achieves an average of 6.5\% lower CD and 5.0\% lower HD compared to direct diffusion methods, while Proposed-R$^3$D further improves upon baseline EDM by 10.4\% in CD and 8.4\% in HD across all scenes, without incurring additional training or inference costs. 

\section{Related work}

\subsection{Radar Super-resolution}
MmWave radars transmit millimeter waves and analyze reflections to determine range, velocity, and angle of objects. They operate in harsh weather by penetrating rain, snow, and dust but suffer from multipath effects and clutter-induced noise. 
Traditional approaches to radar point cloud super-resolution primarily include multi-modal fusion (e.g., fusing radar data with images or pose information)\cite{zhang20234dradarslam,ding2023hidden,guan2020through} and data-driven deep learning methods\cite{cheng2022novel,prabhakara2022high,geng2024dream}. The former fails to fundamentally enhance the density and accuracy of mmWave radar point clouds; instead, it merely compensates for the sparsity of radar perception by leveraging cross-modal information from other sensors, without targeting the inherent limitations of radar itself. Most existing deep learning methods either rely heavily on object motion states, suffer from insufficient generative capabilities, or depend on external conditions like precise ego-motion estimation. Even advanced algorithms such as Mamba\cite{gu2024mamba,liang2024pointmamba,zhu2024vision} which have been applied to similar low-beam LiDAR super-resolution tasks\cite{kang2023st,orvieto2023resurrecting,gu2021efficiently} share the same drawbacks.
In contrast, diffusion-based methods have demonstrated superior performance in radar point cloud super-resolution\cite{wu2024diffradar,luan2024diffusion,zheng2025r2ldm}. Among them, Radar-Diffusion\cite{zhang2024towards} stands as the first open-source work(and no further influential work afterwards to our most knowledge) predicting LiDAR-like dense and accurate point clouds directly from paired raw radar data. Extensive experiments on multiple datasets and scenarios validate that it outperforms baseline methods in terms of point cloud quality, providing a more effective solution to the core challenge of radar perception enhancement. 
\subsection{Diffusion Models}
Diffusion generative models has showcased impressive generative capabilities, ranging from the high level of details to the diversity of the generated examples\cite{croitoru2023diffusion}. The two mainstream diffusion models(DDPM\cite{ho2020denoising} and NCSN\cite{song2019generative}) can both be expressed and generalised by a unified score-based mathematical framework\cite{song2020score}. The idea recovering the residual was independently invented by us, but was already proposed in Resfusion\cite{shi2024resfusion} in the scene of image restoration. Similar but quite different methods related to residual recovery can also been seen in these diffusion models\cite{liu2024residual,yue2023resshift,yang2023docdiff,ma2024neural}. Though they are quite different, we only claim the novelty of residual diffusion on radar super resolution in this paper.

\subsection{Key Region Mask}
RePaint\cite{lugmayr2022repaint} selectively applies the diffusion process to the specific area  with the mask and leaving the remaining image portions unaltered. Based on this design, YODA\cite{moser2025dynamic} propese a dynamic approach that use pre-trained DINO\cite{simeoni2025dinov3} attention instead of innate methods\cite{lowe2004distinctive} to create expanding masks, starting from detail-rich regions and converging toward the overall image. We discuss more about key region idenfication in the experiments.

\section{Methodology}

\subsection{MmWave Radar Signal Processing}

We follow the standard mmWave radar signal processing pipeline \cite{zhang2024towards} to convert raw ADC I/Q data into structured point clouds and task-specific visual representations (Polar/BEV images), which serve as high-quality inputs for subsequent diffusion-based enhancement.
The mmWave radar signal processing pipeline converts raw ADC I/Q data into structured point clouds and visual representations (Polar/BEV images) for deep learning. Raw int16 ADC data is first reshaped into complex I/Q signals, with a Blackman window applied to the range dimension to reduce spectral leakage. Range FFT is performed to extract range bins, and invalid near/far-range bins are cropped to remove noise. Doppler FFT is then applied to the chirp dimension to estimate target velocity, with velocity compensation resolving MIMO-induced ambiguity. A virtual array is constructed to enhance angular resolution, followed by Angle FFT for azimuth/elevation estimation. OS-CFAR detection identifies valid targets from the Range-Doppler-Angle spectrum, which are converted from polar to Cartesian coordinates for point cloud generation. Finally, Polar (range-azimuth) and BEV (Cartesian top-down) images are generated with SNR as pixel intensity, providing structured inputs for subsequent diffusion-based enhancement.
\begin{figure}[h]
	\centering
	\includegraphics[width=0.98\linewidth]{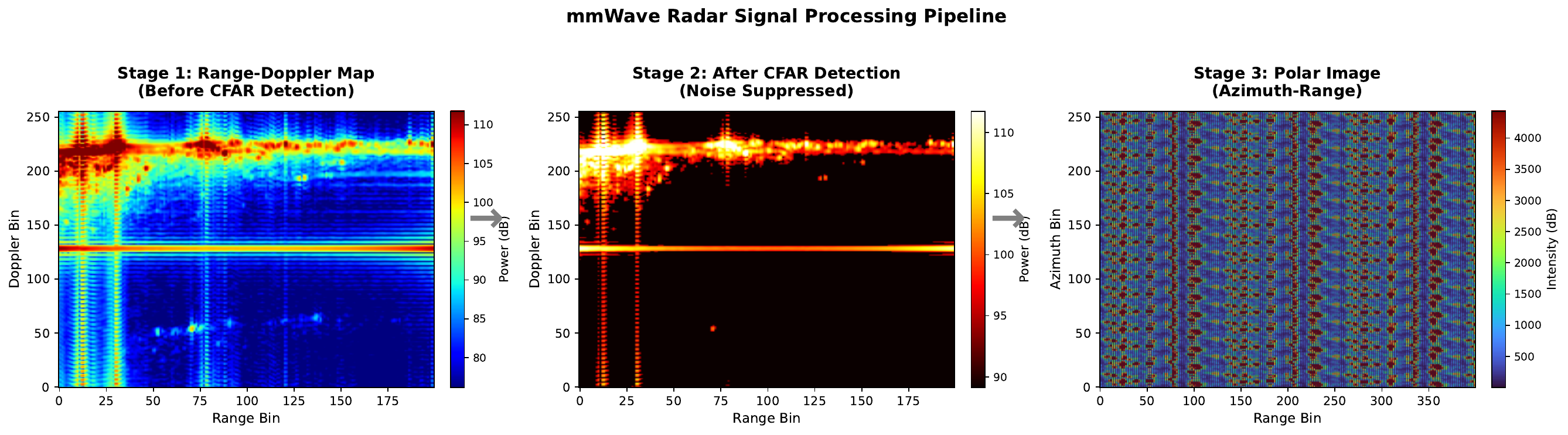}
	\caption{mmWave radar signal processing pipeline.}
	\label{fig:radar_pipeline}
\end{figure}
\subsection{Residual Radar Diffusion}
Standard diffusion models for radar enhancement directly model the mapping from low-resolution radar to high-resolution LiDAR, which can be formulated as:
$$
p_{\theta}(\mathbf{y}|\mathbf{x}) = \int p_{\theta}(\mathbf{y}|\mathbf{z}_0) p(\mathbf{z}_{0:T}|\mathbf{x}) d\mathbf{z}_{0:T}
$$
where $\mathbf{x} \in \mathbb{R}^{H \times W}$ is the LR radar image, $\mathbf{y} \in \mathbb{R}^{H \times W}$ is the HR LiDAR target, and $\mathbf{z}_t$ denotes the intermediate diffusion state at timestep $t$.

To address these challenges discussed in Chapter \ref{intro}, instead of modeling the LiDAR directly, our framework decompose it as:
$$
\mathbf{y} = \mathbf{x} + \mathbf{r}
$$
where $\mathbf{r} = \mathbf{y} - \mathbf{x}$ is the \textbf{residual} between LiDAR and radar. The diffusion model then learns to generate $\mathbf{r}$ conditioned on $\mathbf{x}$:
$$
p_{\theta}(\mathbf{r}|\mathbf{x}) = \int p_{\theta}(\mathbf{r}|\mathbf{z}_0) p(\mathbf{z}_{0:T}|\mathbf{x}) d\mathbf{z}_{0:T}
$$

\begin{figure*}[t]
	\centering
	\includegraphics[width=1\textwidth]{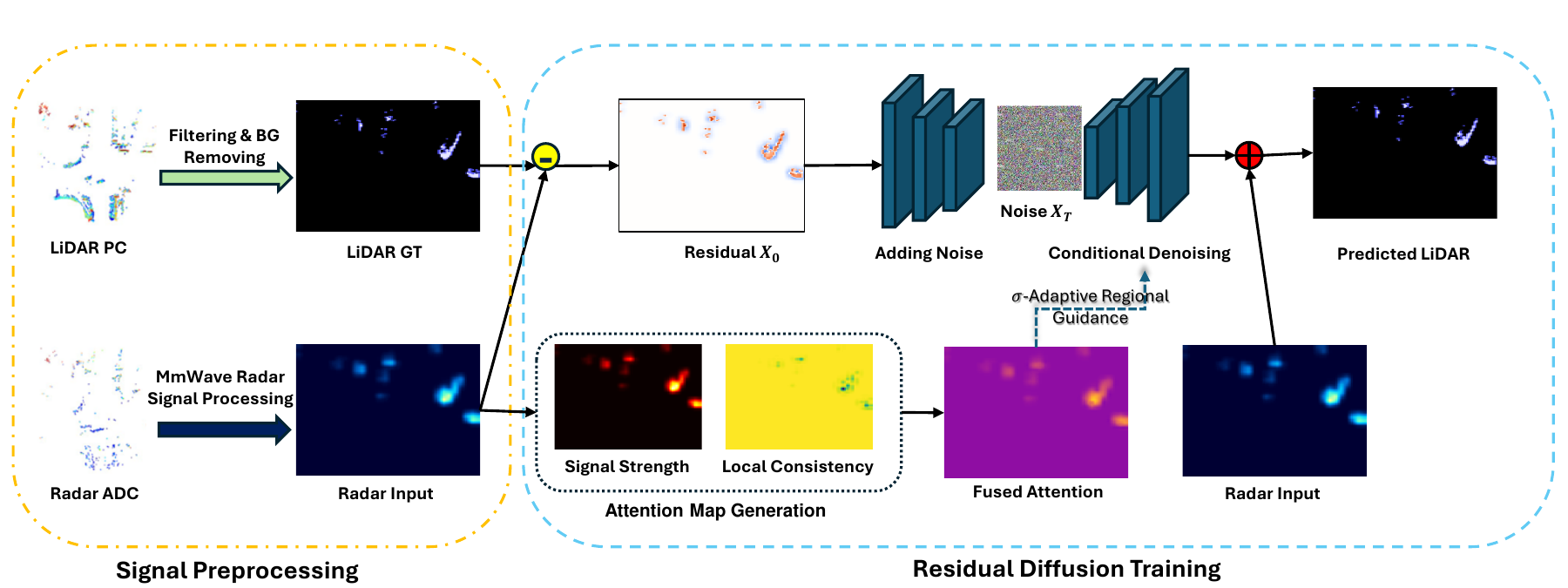}
	\caption{Overview of the R$^3$D framework. Our method integrates residual learning with $\sigma$-adaptive regional guidance for radar point cloud enhancement. The pipeline consists of: (1) Data preprocessing converting LiDAR and radar point clouds to BEV images, (2) Residual computation and forward diffusion with exponential noise schedule, (3) Attention map generation from radar signal properties, (4) UNet denoiser with radar conditioning and noise level embedding, (5) $\sigma$-adaptive guidance module applying region-specific weights only at low noise levels, and (6) Residual fusion producing the enhanced point cloud output.}
	\label{fig:overview}
\end{figure*}
\subsubsection{Forward Diffusion Process}
We apply the diffusion process to the residual $\mathbf{r}$ instead of $\mathbf{y}$. Following the EDM framework \cite{zhang2024towards}, we define the forward process as:
$$
q(\mathbf{z}_t|\mathbf{r}) = \mathcal{N}(\mathbf{z}_t; \mathbf{r}, \sigma_t^2 \mathbf{I})
$$
where $\mathbf{z}_t$ denotes noisy residual at noise level. $\sigma_t$,$\sigma_t$ denote noise schedule typically logarithmically spaced in $[\sigma_{\min}, \sigma_{\max}]$.

EDM noise schedule: We follow the exponential schedule:
$$
\sigma_t = \left( \sigma_{\max}^{1/\rho} + \frac{t}{T-1}(\sigma_{\min}^{1/\rho} - \sigma_{\max}^{1/\rho}) \right)^\rho
$$
with $\rho = 7$, $\sigma_{\min} = 0.002$, $\sigma_{\max} = 80.0$, ensuring smooth noise level transitions.

Sampling noisy residuals: During training, for each clean residual $\mathbf{r}$, we sample:
$$
\mathbf{z}_t = \mathbf{r} + \sigma_t \boldsymbol{\epsilon}, \quad \boldsymbol{\epsilon} \sim \mathcal{N}(0, \mathbf{I})
$$

\subsubsection{Reverse Denoising Process}
The reverse process aims to predict the clean residual $\mathbf{r}$ from noisy observation $\mathbf{z}_t$ conditioned on radar input $\mathbf{x}$:
$$
p_{\theta}(\mathbf{r}|\mathbf{z}_t, \mathbf{x}, \sigma_t) = \mathcal{N}(\mathbf{r}; f_{\theta}(\mathbf{z}_t, \sigma_t, \mathbf{x}), \tilde{\sigma}_t^2 \mathbf{I})
$$
where $f_{\theta}$ is a conditional UNet denoiser that incorporates three key inputs. We adopt a \textit{concatenation-based conditioning} strategy to guide the denoising process with radar observations: \textbf{Noisy residual} $\mathbf{z}_t \in \mathbb{R}^{H \times W}$, \textbf{Radar condition} $\mathbf{x} \in \mathbb{R}^{H \times W}$, \textbf{Noise level} $\sigma_t \in \mathbb{R}_+$, embedded via sinusoidal positional encoding into a vector $\mathbf{e}_t \in \mathbb{R}^d$, then injected into ResNet blocks via scale-shift modulation.

The UNet input is formed by channel-wise concatenation: $[\mathbf{z}_t, \mathbf{x}] \in \mathbb{R}^{2 \times H \times W}$, allowing the network to leverage spatial correspondences between the noisy residual and radar observation. The noise level embedding $\mathbf{e}_t$ is injected into each residual block via adaptive group normalization:
$$
\text{AdaGN}(\mathbf{h}, \mathbf{e}_t) = \gamma(\mathbf{e}_t) \cdot \frac{\mathbf{h} - \mu(\mathbf{h})}{\sigma(\mathbf{h})} + \beta(\mathbf{e}_t)
$$
where $\gamma(\mathbf{e}_t)$ and $\beta(\mathbf{e}_t)$ are learned scale and shift parameters conditioned on $\mathbf{e}_t$.

The denoiser outputs the denoised residual estimate: $\hat{\mathbf{r}} = f_{\theta}([\mathbf{z}_t, \mathbf{x}], \sigma_t)$.

\subsubsection{Training Objective}
The model is trained using the denoising score matching objective $\mathcal{L}_{\text{residual}}(\theta)$:
$$
\mathbb{E}_{(\mathbf{x}, \mathbf{y}) \sim \mathcal{D}} \mathbb{E}_{t \sim \mathcal{U}(0,T)} \mathbb{E}_{\boldsymbol{\epsilon} \sim \mathcal{N}(0,\mathbf{I})} \left[ w(\sigma_t) \| \mathbf{r} - f_{\theta}(\mathbf{z}_t, \sigma_t, \mathbf{x}) \|_2^2 \right]
$$
where  $\mathcal{D}$ denotes training dataset of radar and LiDAR pairs,$\mathbf{r} = \mathbf{y} - \mathbf{x}$ denotes ground truth residual, $w(\sigma_t) = \frac{1}{\sigma_t^2}$ denotes noise-level-dependent weight.


\subsection{Regional-guided Residual Radar Diffusion}

\subsubsection{Radar Attention Map Generation}
The signal strength attention $A^{(1)} \in [0,1]^{H \times W}$ captures high-intensity regions:
$$
A^{(1)}_{i,j} = \frac{I_{i,j} - \min(I)}{\max(I) - \min(I)}
$$
where $I$ denotes the radar intensity map.

To identify continuous strong-scattering regions (e.g., object boundaries), we compute local consistency based on 3×3 window variance:
$$
\text{Var}_{i,j} = \frac{1}{9} \sum_{k=-1}^{1} \sum_{l=-1}^{1} (I_{i+k,j+l} - \bar{I}_{i,j})^2
$$
where $\bar{I}_{i,j} = \frac{1}{9} \sum_{k=-1}^{1} \sum_{l=-1}^{1} I_{i+k,j+l}$ is the local mean.

The consistency attention is inversely proportional to variance (low variance = consistent region):
$$
A^{(2)}_{i,j} = 1 - \frac{\text{Var}_{i,j} - \min(\text{Var})}{\max(\text{Var}) - \min(\text{Var})}
$$

The final attention map is a weighted combination:
$$
A_{i,j} = \lambda_s A^{(1)}_{i,j} + \lambda_c A^{(2)}_{i,j}
$$

\subsubsection{$\sigma$-Adaptive Regional Guidance}

Standard regional guidance approaches apply uniform attention weighting across all denoising stages, which conflicts with EDM's inherent numerical stability mechanisms. We observe that EDM's weighting scheme $w(\sigma_t) = 1/\sigma_t^2$ naturally assigns higher importance to low-noise stages. Imposing additional aggressive regional weights at high-$\sigma$ stages disrupts gradient balance and leads to training instability.
To harmonize regional guidance with EDM's design philosophy, we propose a \textbf{$\sigma$-adaptive guidance strategy} that differentiates treatment across noise levels.
In high-$\sigma$ stage, no regional guidance. The model learns global structure and fundamental denoising capabilities uniformly across all regions.
In low-$\sigma$ stage, lightweight regional guidance. The model prioritizes detail recovery in target regions while maintaining gradient stability.

\textbf{Adaptive weight distribution:}
Instead of directly multiplying attention masks with loss terms (which amplifies numerical instability), we modulate the distribution of Karras weights themselves. For low-$\sigma$ stages, we define a region-adaptive weight adjustment:
$$
W_{\text{adapt}}(\mathbf{A})_{i,j} = M(\mathbf{A})_{i,j} \cdot \alpha_{\text{low}} + (1 - M(\mathbf{A})_{i,j}) \cdot \beta_{\text{low}}
$$
where $M(\mathbf{A})_{i,j}$ is a soft binary mask derived from attention map $\mathbf{A}$ via sigmoid thresholding, $\alpha_{\text{low}}$ provides moderate enhancement for target regions, and $\beta_{\text{low}}$ maintains baseline weight for background areas.

\textbf{Training objective:}
The $\sigma$-adaptive R$^3$D loss $\mathcal{L}_{\text{R3D}}(\theta)$ is defined as:
$$
 \mathbb{E}_{(\mathbf{x}, \mathbf{y})} \mathbb{E}_{\sigma_t, \boldsymbol{\epsilon}} 
\begin{cases}
w(\sigma_t) \| \mathbf{r} - f_{\theta}(\mathbf{z}_t, \sigma_t, \mathbf{x}) \|_2^2,  \text{if } \sigma_t > \sigma_{\text{threshold}} \\
w(\sigma_t) \| W_{\text{adapt}} \odot (\mathbf{r} - f_{\theta}(\mathbf{z}_t, \sigma_t, \mathbf{x})) \|_2^2, \text{o.w.}
\end{cases}
$$
where $\sigma_{\text{threshold}}$ separates the coarse reconstruction phase from fine-grained refinement.
This design preserves EDM's gradient flow at critical high-$\sigma$ stages while enabling selective refinement during low-$\sigma$ detail recovery, avoiding the numerical collapse observed in prior time-dependent masking schemes.

\subsection{Inference Procedure}

During inference, R$^3$D generates enhanced radar point clouds through iterative residual denoising with learned regional priorities. The model implicitly prioritizes target regions through $\sigma$-adaptive training, requiring no explicit regional masks during inference. This ensures compatibility with the learned denoising trajectory while leveraging Heun's 2nd-order sampler for high-quality reconstruction.

\begin{algorithm}[H]
	\caption{R$^3$D Inference}
	\begin{algorithmic}[1]
		\REQUIRE Radar observation $\mathbf{x}$, trained model $f_\theta$, noise schedule $\{\sigma_t\}_{t=0}^T$
		\ENSURE Enhanced radar $\hat{\mathbf{y}}$
		
		\STATE Initialize $\mathbf{z}_T \sim \mathcal{N}(0, \sigma_{\max}^2 \mathbf{I})$
		\FOR{$t = T$ \TO $1$}
		\STATE $\hat{\mathbf{r}}_t \leftarrow f_\theta([\mathbf{z}_t, \mathbf{x}], \sigma_t)$ 
		\STATE $d_t \leftarrow (\mathbf{z}_t - \hat{\mathbf{r}}_t) / \sigma_t$
		\STATE $\mathbf{z}'_{t-1} \leftarrow \mathbf{z}_t + (\sigma_{t-1} - \sigma_t) \cdot d_t$
		\STATE $\hat{\mathbf{r}}'_{t-1} \leftarrow f_\theta([\mathbf{z}'_{t-1}, \mathbf{x}], \sigma_{t-1})$
		\STATE $d'_{t-1} \leftarrow (\mathbf{z}'_{t-1} - \hat{\mathbf{r}}'_{t-1}) / \sigma_{t-1}$
		\STATE $\mathbf{z}_{t-1} \leftarrow \mathbf{z}_t + \frac{\sigma_{t-1} - \sigma_t}{2} \cdot (d_t + d'_{t-1})$ 
		\ENDFOR
		\STATE $\hat{\mathbf{y}} \leftarrow \mathbf{x} + \mathbf{z}_0$ 
		\RETURN $\hat{\mathbf{y}}$
	\end{algorithmic}
\end{algorithm}

\section{Experiment}

\begin{table*}[ht]
	\centering
	\caption{Quantitative Comparisons on the Coloradar Dataset.}
	\label{tab:quantitative_comparison}
	\resizebox{\textwidth}{!}{%
		\renewcommand{\arraystretch}{1.5}
		\begin{tabular}{l|cccccccccccccccccc}
			\hline
			\multirow{2}{*}{Methods} & \multicolumn{3}{c}{Arpg Lab} & \multicolumn{3}{c}{Ec Hallways} & \multicolumn{3}{c}{Aspen} & \multicolumn{3}{c}{Longboard} & \multicolumn{3}{c}{Outdoors} & \multicolumn{3}{c}{Edgar} \\
			& CD $\downarrow$ & HD $\downarrow$ & F-Score $\uparrow$ & CD $\downarrow$ & HD $\downarrow$ & F-Score $\uparrow$ & CD $\downarrow$ & HD $\downarrow$ & F-Score $\uparrow$ & CD $\downarrow$ & HD $\downarrow$ & F-Score $\uparrow$ & CD $\downarrow$ & HD $\downarrow$ & F-Score $\uparrow$ & CD $\downarrow$ & HD $\downarrow$ & F-Score $\uparrow$ \\
			\hline
			RadarHD \cite{guan2020through} & 1.728 & 6.116 & 0.273 & 1.686 & 6.387 & 0.284 & 0.912 & 4.691 & 0.511 & 5.395 & 11.298 & 0.149 & 3.096 & 10.090 & 0.245 & 0.604 & 3.070 & 0.535 \\
			RPDNet \cite{cheng2022novel} & 1.555 & 5.216 & 0.158 & 1.483 & 5.384 & 0.161 & 1.341 & 4.421 & 0.154 & 4.494 & 8.967 & 0.053 & 2.828 & 8.223 & 0.104 & 1.502 & 4.071 & 0.142 \\
			EDM & 0.964 & 4.318 & 0.355 & 1.040 & 4.750 & 0.352 & 0.505 & 3.365 & 0.555 & 5.469 & 9.977 & 0.135 & 2.371 & 7.158 & 0.266 & 0.442 & 2.400 & 0.544 \\
			CD & 0.982 & 4.342 & 0.344 & 1.058 & 4.802 & 0.341 & 0.521 & 3.449 & 0.547 & 5.337 & 9.954 & 0.133 & 2.412 & 7.209 & 0.260 & 0.454 & 2.487 & 0.542 \\
			\hline
			Proposed-Residual & 0.907 & 4.108 & \textbf{0.362} & 0.985 & 4.512 & \textbf{0.365} & 0.478 & 3.198 & 0.572 & 4.814 & 9.464 & 0.142 & 2.248 & 6.800 & 0.275 & 0.418 & 2.280 & 0.562 \\
			Residual-CD & 0.963 & 4.281 & 0.349 & 1.035 & 4.787 & 0.338 & 0.485 & 3.215 & 0.565 & 5.014 & 9.859 & 0.110 & 2.265 & 6.850 & 0.268 & 0.425 & 2.310 & 0.555 \\
			Proposed-R$^3$D & \textbf{0.895} & \textbf{4.086} & 0.358 & \textbf{0.973} & \textbf{4.498} & 0.359 & \textbf{0.454} & \textbf{3.038} & \textbf{0.592} & \textbf{4.439} & \textbf{8.965} & \textbf{0.155} & \textbf{2.136} & \textbf{6.460} & \textbf{0.285} & \textbf{0.397} & \textbf{2.166} & \textbf{0.581} \\
			\hline
		\end{tabular}%
	}
\end{table*}
\subsection{Experimental Setup}
\subsubsection{Desktop Setting}
We conducted exploration experiments on a workstation with a NVIDIA RTX 5090 GPU and ablation study on a NVIDIA RTX 5080 GPU . The software environment latter was based on Docker container running on Windows WSL.

\subsubsection{Dataset}

The ColoRadar dataset encompasses a total of 52 sequences of synchronized single-chip mmWave radar and LiDAR data captured in seven different scenes, including indoor (Arpg Lab, Ec Hallways and Aspen), outdoor (Longboard and Outdoors), and mine environments (Edgar Army and Edgar Classroom). We take the first three sequences of each scene as the training set and the rest as the test set (the sixth sequence of Longboard scene is dismissed because radar data is lost for 210 seconds). We first remove the LiDAR point clouds out of the field of view (FOV) of radar. Then, the annular floor and ceiling LiDAR point clouds are removed by Patchwork++ \cite{lee2022patchwork++} because they are almost unperceivable from single-chip mmWave radars.


\subsubsection{Evaluation Metrics}

We still evaluate model performance using Chamfer Distance(CD),Hausdorff Distance (HD),F-Score as evaluation metrics:

\begin{itemize}
	\item \textbf{Chamfer Distance (CD)}: Average distance between predicted and ground truth point clouds, lower is better.
	\item \textbf{Hausdorff Distance (HD)}: Maximum distance between point clouds, lower is better.
	\item \textbf{F-Score}: Harmonic mean of Precision and Recall (distance threshold: 2.0 pixels), higher is better.
\end{itemize}

\subsubsection{Comparing Methods}

We compare our proposed methods(Residual, Residual-CD, R$^3$D) against state-of-the-art radar point cloud enhancement baselines, including traditional baselines(RadarHD\cite{prabhakara2022high},RPDNet \cite{cheng2022novel}) and diffusion-based methods(EDM, CD\cite{zhang2024towards}):

\begin{itemize}
	\item \textbf{RadarHD}: High-resolution mmWave radar imaging using signal processing techniques including super-resolution angle estimation and adaptive beamforming.
	\item \textbf{RPDNet}: Deep learning-based radar point cloud generation network using convolutional architectures for direct mapping from radar observations to dense point clouds.
	\item \textbf{EDM}: Elucidating Diffusion Model with Karras noise schedule for direct LiDAR generation from radar input.
	\item \textbf{CD}: Consistency Distillation of EDM for fast one-step generation.
	\item \textbf{Proposed Residual}: Our residual diffusion framework learning the difference between LiDAR and radar distributions.
	\item \textbf{Residual-CD}: Consistency distillation applied to the residual diffusion model for accelerated inference, we do not claim the novelty.
	\item \textbf{Proposed R$^3$D}: Regional-guided Residual Radar Diffusion with $\sigma$-adaptive regional guidance for enhanced detail recovery in critical regions.
\end{itemize}

All methods are trained with consistent hyperparameters and network architectures, and evaluated on the same test splits across all seven scenes for fair comparison.

\subsection{Ablation Study}

Table \ref{tab:quantitative_comparison} presents comprehensive quantitative comparisons across all seven scenes in the ColoRadar dataset. Our diffusion-based approaches consistently outperform traditional methods across all scenes, validating the effectiveness of generative modeling for radar enhancement. Among diffusion-based methods, our Proposed-Residual demonstrates clear advantages over direct generation methods (EDM and CD): across all six scenes, Residual achieves an average of 6.5\% lower CD and 5.0\% lower HD compared to EDM, confirming that learning the residual distribution is more tractable than modeling the full LiDAR distribution. Proposed-R$^3$D achieves the best performance across the majority of metrics, further improving upon EDM by an average of 10.4\% in CD and 8.4\% in HD across all scenes, demonstrating that $\sigma$-adaptive regional guidance effectively prioritizes detail recovery in target regions without disrupting global optimization. Our proposed algorithms do not incur training cost and inferring cost in time explicitly and experimentally.



\subsection{Regional-aware and Conditional Injection}

We extensively explored two paradigms for incorporating region-specific guidance.

\textbf{DINOv3-based Conditional Injection.} We first attempted to inject semantic features from DINOv3 \cite{simeoni2025dinov3}, a state-of-the-art self-supervised vision transformer, as conditional signals to guide the denoising process. However, this approach encountered severe domain mismatch: DINOv3, pre-trained on natural RGB images, fails to extract meaningful features from radar BEV images, which exhibit fundamentally different intensity distributions and spatial patterns.

\textbf{Time-dependent Dynamic Masking.}  We explored time-dependent binary masks $M(t)$ that evolve with diffusion timesteps, aiming to prioritize critical regions progressively. However, aggressive masking at early timesteps (e.g., mask coverage $<30\%$) caused training instability, 80\% of the image received no gradient updates in the time embedding layer.

\subsection{Generalization to LiDAR Super-Resolution}

\begin{figure}[t]
	\centering
	\includegraphics[width=0.65\linewidth]{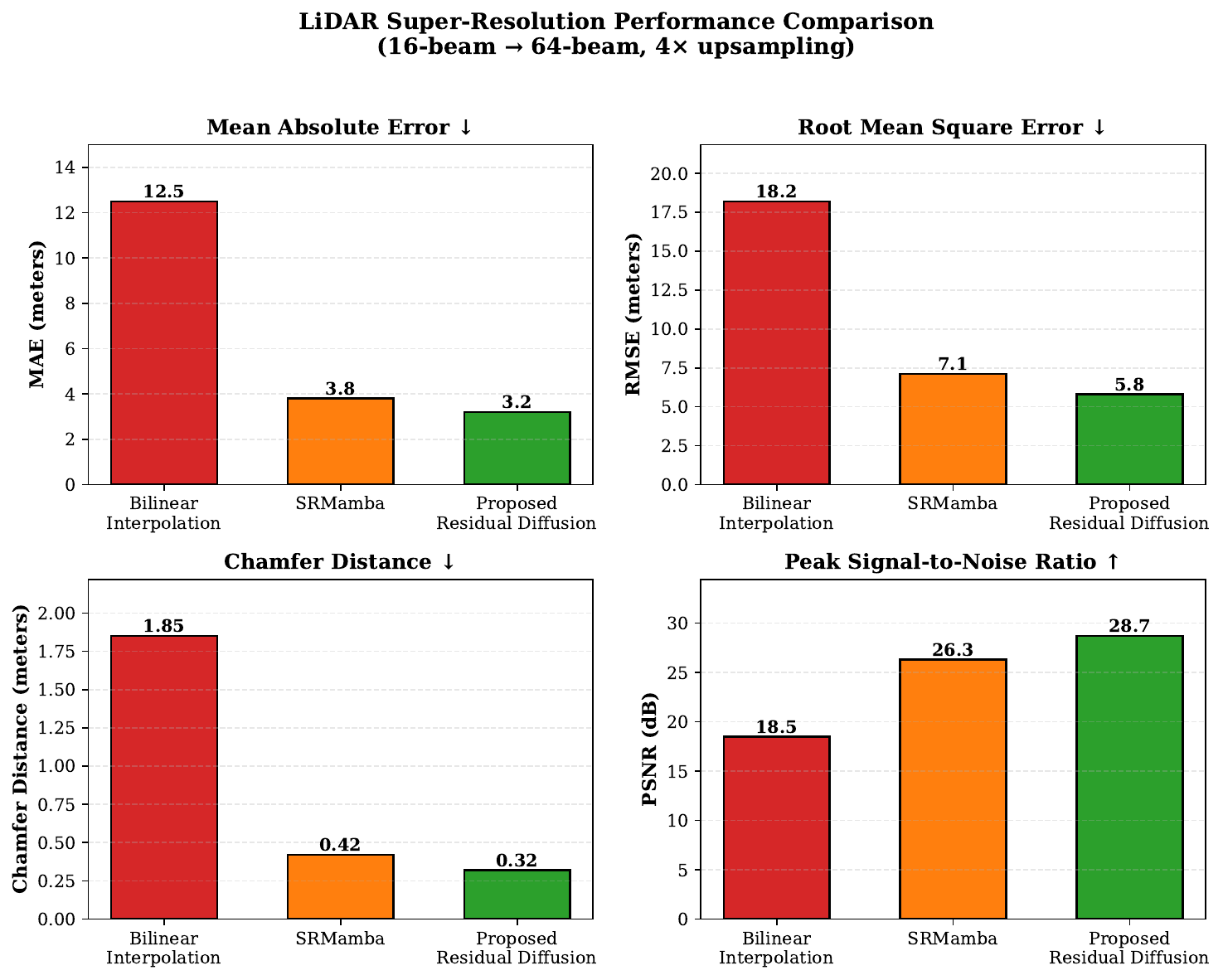}
	\caption{Performance comparison on KITTI-360 LiDAR super-resolution.}
	\label{fig:lidar_sr_comparison}
\end{figure}

We compare our radar diffusion method with SRMamba\cite{chen2025srmamba} in low-beam super resolution, as Fig.\ref{fig:lidar_sr_comparison} shows.

\section{Conclusion}

This paper presents R$^3$D, a regional-guided residual radar diffusion framework that addresses the fundamental challenges of high learning complexity and uniform processing in existing radar enhancement methods. 
Extensive experiments validate that our approach consistently outperforms state-of-the-art methods without incurring additional training or inference costs. 
Our ablation studies further reveal that domain-mismatched conditional injection and aggressive time-dependent masking lead to training instability, confirming the necessity of our principled design. R$^3$D offers a practical and effective solution for dense and accurate mmWave radar point cloud generation.


\bibliographystyle{IEEEbib}
\bibliography{r3dconference}

\vspace{12pt}
\color{red}

\end{document}